\documentstyle[12pt]{article}

\catcode`@=11
\newif\if@defeqnsw \@defeqnswtrue

\def\eqnarray{\stepcounter{equation}\let\@currentlabel=\theequation
\if@defeqnsw\global\@eqnswtrue\else\global\@eqnswfalse\fi
\global\@eqnswtrue
\tabskip\@centering\let\\=\@eqncr
$$\halign to \displaywidth\bgroup\hfil\global\@eqcnt\z@
  $\displaystyle\tabskip\z@{##}$&\global\@eqcnt\@ne
  \hfil$\displaystyle{{}##{}}$\hfil
  &\global\@eqcnt\tw@ $\displaystyle{##}$\hfil
  \tabskip\@centering&\llap{##}\tabskip\z@\cr}

\def\yesnumber{\global\@eqnswtrue}

\def\@@eqncr{\let\@tempa\relax\global\advance\@eqcnt by \@ne
    \ifcase\@eqcnt \def\@tempa{& & & &}\or \def\@tempa{& & &}\or
     \def\@tempa{& &}\or \def\@tempa{&}\else\fi
     \@tempa \if@eqnsw\@eqnnum\stepcounter{equation}\fi
     \if@defeqnsw\global\@eqnswtrue\else\global\@eqnswfalse\fi
     \global\@eqcnt\z@\cr}

\def\@eqnacr{{\ifnum0=`}\fi\@ifstar{\@yeqnacr}{\@yeqnacr}}

\def\@yeqnacr{\@ifnextchar [{\@xeqnacr}{\@xeqnacr[\z@]}}

\def\@xeqnacr[#1]{\ifnum0=`{\fi}\cr \noalign{\vskip\jot\vskip #1\relax}}

\def\eqalign{\null\,\vcenter\bgroup\openup1\jot \m@th \let\\=\@eqnacr
\ialign\bgroup\strut
\hfil$\displaystyle{##}$&$\displaystyle{{}##}$\hfil\crcr}
\def\endeqalign{\crcr\egroup\egroup\,}

\def\cases{\left\{\,\vcenter\bgroup\normalbaselines\m@th \let\\=\@eqnacr
    \ialign\bgroup$##\hfil$&\quad##\hfil\crcr}
\def\endcases{\crcr\egroup\egroup\right.}

\def\eqalignno{\stepcounter{equation}\let\@currentlabel=\theequation
\if@defeqnsw\global\@eqnswtrue\else\global\@eqnswfalse\fi
\let\\=\@eqncr
$$\displ@y \tabskip\@centering \halign to \displaywidth\bgroup
  \global\@eqcnt\@ne\hfil
  $\@lign\displaystyle{##}$\tabskip\z@skip&\global\@eqcnt\tw@
  $\@lign\displaystyle{{}##}$\hfil\tabskip\@centering&
  \llap{\@lign##}\tabskip\z@skip\crcr}

\def\endeqalignno{\@@eqncr\egroup
      \global\advance\c@equation\m@ne$$\global\@ignoretrue}

\@namedef{eqalignno*}{\@defeqnswfalse\eqalignno}
\@namedef{endeqalignno*}{\endeqalignno}

\def\eqaligntwo{\stepcounter{equation}\let\@currentlabel=\theequation
\if@defeqnsw\global\@eqnswtrue\else\global\@eqnswfalse\fi
\let\\=\@eqncr
$$\displ@y \tabskip\@centering \halign to \displaywidth\bgroup
  \global\@eqcnt\m@ne\hfil
  $\@lign\displaystyle{##}$\tabskip\z@skip&\global\@eqcnt\z@
  $\@lign\displaystyle{{}##}$\hfil\qquad&\global\@eqcnt\@ne
  \hfil$\@lign\displaystyle{##}$&\global\@eqcnt\tw@
  $\@lign\displaystyle{{}##}$\hfil\tabskip\@centering&
  \llap{\@lign##}\tabskip\z@skip\crcr}

\def\endeqaligntwo{\@@eqncr\egroup
      \global\advance\c@equation\m@ne$$\global\@ignoretrue}

\@namedef{eqaligntwo*}{\@defeqnswfalse\eqaligntwo}
\@namedef{endeqaligntwo*}{\endeqaligntwo}

\newtoks\@stequation

\def\subequations{\refstepcounter{equation}%
  \edef\@savedequation{\the\c@equation}%
  \@stequation=\expandafter{\theequation}
  \edef\@savedtheequation{\the\@stequation}
  \edef\oldtheequation{\theequation}%
  \setcounter{equation}{0}%
  \def\theequation{\oldtheequation\alph{equation}}}

\def\endsubequations{%
  \setcounter{equation}{\@savedequation}%
  \@stequation=\expandafter{\@savedtheequation}%
  \edef\theequation{\the\@stequation}%
  \global\@ignoretrue}

\def\big#1{{\hbox{$\left#1\vcenter to1.428\ht\strutbox{}\right.\n@space$}}}
\def\Big#1{{\hbox{$\left#1\vcenter to2.142\ht\strutbox{}\right.\n@space$}}}
\def\bigg#1{{\hbox{$\left#1\vcenter to2.857\ht\strutbox{}\right.\n@space$}}}
\def\Bigg#1{{\hbox{$\left#1\vcenter to3.571\ht\strutbox{}\right.\n@space$}}}

\catcode`@=12

\begin{document}

\newcommand{\as}{\mbox{$\alpha_S$}}
\newcommand{\msqav}{\mbox{$\langle m_{\tilde{q}} \rangle$}}
\newcommand{\be}{\begin{equation}}
\newcommand{\ee}{\end{equation}}
\newcommand{\een}{\end{subequations}}
\newcommand{\ben}{\begin{subequations}}
\newcommand{\beq}{\begin{eqalignno}}
\newcommand{\eeq}{\end{eqalignno}}
\newcommand{\tanb}{\mbox{$\tan \! \beta$}}
\newcommand{\eplem}{\mbox{$e^+e^-$}}
\newcommand{\rs}{\mbox{$\sqrt{s}$}}
\newcommand{\s}{\\ \vspace*{-3mm} }
\newcommand{\non}{\nonumber}
\newcommand{\nn}{\noindent}
\newcommand{\mt}{m_q}
\newcommand{\msq}{\mbox{$m_{\tilde{q}}$}}
\newcommand{\msi}{\mbox{$m_{\tilde{q}_i}$}}
\newcommand{\msj}{\mbox{$m_{\tilde{q}_j}$}}
\newcommand{\mso}{\mbox{$m_{\tilde{q}_1}$}}
\newcommand{\mst}{\mbox{$m_{\tilde{q}_2}$}}
\newcommand{\mg}{\mbox{$m_{\tilde{g}}$}}
\newcommand{\glgl}{\mbox{$\tilde{g} \tilde{g}$}}
\newcommand{\hgl}{\mbox{$h \rightarrow \glgl$}}
\newcommand{\Pgl}{\mbox{$P \rightarrow \glgl$}}
\newcommand{\Zgl}{\mbox{$Z \rightarrow \glgl$}}
\newcommand{\is}{2I_q^{3L}}
\newcommand{\heta}{\theta_q}
\newcommand{\bbbar}{\mbox{$b \bar{b}$}}
\newcommand{\ttbar}{\mbox{$t \bar{t}$}}
\newcommand{\ccbar}{\mbox{$c \bar{c}$}}
\newcommand{\qqbar}{\mbox{$q \bar{q}$}}
\newcommand{\bq}{\beta_{\tilde{g}}}
\newcommand{\ct}{\cos^2\theta}
\newcommand{\st}{\sin^2\theta}
\newcommand{\sg}{\tilde{g}}
\newcommand{\sgg}{\tilde{g} \tilde{g}}
\renewcommand{\thefootnote}{\fnsymbol{footnote} }

\pagestyle{empty}

\font\fortssbx=cmssbx10 scaled \magstep2
\hbox to \hsize{
\hskip.5in \raise.1in\hbox{\fortssbx University of Wisconsin - Madison}
\hfill$\vcenter{\hbox{\bf MADPH--94--853}
\hbox{\bf UdeM-GPP--94--08}
            \hbox{October 1994}}$ }

\vspace*{1cm}
\begin{center}
{\Large \bf Higgs and Z boson decays into light gluinos}

\vspace*{1cm}

{\sc A.~Djouadi}$^1\footnote{NSERC fellow}$ and {\sc M.~Drees}$^2
$\footnote{Heisenberg fellow}

\vspace*{.6cm}

\rm{$^1$ Groupe de Physique des Particules, Universit\'e de Montr\'eal,
Case 6128A, }

\vspace*{.1cm}

\rm{H3C 3J7 Montr\'eal PQ, Canada}

\vspace*{.5cm}

\rm{$^2$ Physics Department, University of Wisconsin, Madison, WI 53706,
USA}

\end{center}
\vspace*{1cm}

\begin{abstract}

\nn We calculate the decay rate of scalar and pseudoscalar Higgs bosons into a
pair of gluinos, within the Minimal Supersymmetric Standard Model. In the
theoretically and experimentally allowed light gluino window, $\mg \sim $ 3--5
GeV, gluino pairs can completely dominate the decays of the light scalar Higgs
boson and play a prominent role in the decay of the pseudoscalar Higgs
boson. This would alter the limits obtained from $Z$ decays on the lightest
CP--even and CP--odd Higgs bosons, and could jeopardize the search for these
Higgs particles at future hadron colliders. In contrast, the branching ratio
for the two--body decay of $Z$ bosons into pairs of light gluinos is less than
0.1\%.
\end{abstract}

\newpage

\pagestyle{plain}
\renewcommand{\thefootnote}{\arabic{footnote} }
\setcounter{footnote}{0}

\subsection*{1. Introduction}

The last couple of years have seen renewed interest \cite{1} in the possible
existence of very light gluinos, with mass $\mg \leq 5$ GeV. A recent analysis
by Farrar \cite{farrar} concludes that only the upper end of this range is
still open, roughly 3 GeV $\leq \mg \leq$ 5 GeV. However, even this window
would be sufficient to allow for a substantial modification of the running of
\as, which was the original motivation \cite{1} for this scenario. Moreover,
since the decays of light gluinos might only produce a small amount of missing
energy \cite{3}, bounds on squark masses derived from searches \cite{pp} for
events with large missing $E_T$ at $p \overline{p}$ colliders are not valid in
this scenario. \s

In this paper we point out that the existence of light gluinos could also
substantially complicate the search for the Higgs bosons predicted by
supersymmetric models. To be specific, we work within the framework of the
minimal supersymmetric standard model or MSSM, which contains two $SU(2)$
doublets of Higgs superfields \cite{mssm,rev}. We will not assume unification
of
gaugino masses, which would impose severe constraints on models with light
gluinos \cite{valle}, nor do we require radiative breaking of the $SU(2)$
symmetry, which appears to be inconsistent with the existence of light
gluinos, at least in the minimal model \cite{diaz}. We find that the \glgl\
final state can completely dominate the decay of the light scalar Higgs boson;
in some cases the branching ratios for the usually dominant $b \overline{b}$
and $\tau^+ \tau^-$ final states are reduced by a factor of 1,000 or more.
Light gluinos might also be produced in 70\% of all decays of the pseudoscalar
Higgs boson of the MSSM. This has obvious consequences for Higgs searches that
rely on $b$ or $\tau$ tagging. The decays of supersymmetric Higgs bosons to
gluino pairs have been discussed in the past in ref.~\cite{ng}. However, the
possibility that the gluinos are light and the fact that the mixing between
the two supersymmetric scalar partners of the top quark (which, as we will see
later, is the most important feature in this decay) might be large have not
been considered. Therefore the Higgs $\rightarrow  \tilde{g} \tilde{g}$
branching ratios obtained in ref.~\cite{ng} were very small. \s

It has been known for some time \cite{7} that light gluinos can also be
produced in two--body decays of $Z$ bosons; this decay has recently been
studied in ref.~\cite{norway}. We re--compute the corresponding partial width,
including the effect of mixing between $SU(2)$ doublet and singlet squarks.
Our numerical results agree with those of ref.~\cite{norway}, but our final
expression is much more compact. We find a maximal $Br(\Zgl) \simeq 6 \cdot
10^{-4}$ within our assumptions; such a small branching ratio can only be
probed by a dedicated search for \glgl\ final states. Unfortunately the
failure to detect $\Zgl$ decays would not exclude the possibility that gluino
pairs contribute significantly or dominantly to Higgs boson decays.\s

The remainder of this article is organized as follows. In sec.~2 expressions
for the partial widths for the decays of Higgs and $Z$ bosons into pairs of
light (effectively massless) gluinos are given. In sec.~3 we present numerical
results for the separate branching ratios and discuss possible correlations
between them. Sec. 4 contains a brief summary and conclusions. The Appendix
collects expressions for the decays of Higgs bosons into massive gluinos, as
well as the necessary two-- and three--point functions.

\subsection*{2. Formalism}

The diagrams contributing to the one--loop induced decays of Higgs and $Z$
bosons into a pair of gluinos are shown in  Fig.~1. Since the gluinos are of
Majorana nature and hence identical fermions, one has to antisymmetrize the
decay amplitude; this is achieved by adding the contributions of the two
diagrams where the momenta of the gluinos are interchanged and by multiplying
these contributions by an overall minus sign. In the case of Higgs bosons,
this reduces to just multiplying the amplitudes of the diagrams shown in
Fig.~1 by a factor of two; for the $Z$ boson one needs in addition
to discard the terms that are not proportional to $\gamma_5$. \s

In Fig.~1 one has to include the contributions of both squarks of a given
flavor. As well known \cite{rudaz},
the supersymmetric partners of left-- and right--handed massive quarks mix. The
mass eigenstates $\tilde{q}_1$ and $\tilde{q}_2$ are related to the current
eigenstates $\tilde{q}_L$ and $\tilde{q}_R$ by
\begin{eqnarray}\label{e3p}
\tilde{q}_1=\tilde{q}_L \cos \heta +\tilde{q}_R \sin \heta \ \ , \hspace*{1cm}
\tilde{q}_2=- \tilde{q}_L \sin \heta +\tilde{q}_R \cos \heta.
\end{eqnarray}

The mixing angle $\heta$ as well as the masses \mso, \mst\ of the physical
squarks can be calculated from the following mass matrices\footnote{We ignore
generation mixing between squarks, which in case of the MSSM is only induced
radiatively by weak interactions.}, written in the convention of
ref.~\cite{dn1}:
\ben \label{e4} \beq
{\cal M}^2_{\tilde t} &= \mbox{$ \left( \begin{array}{cc}
m^2_{\tilde{t}_L} + m_t^2 + 0.35 D_Z & - m_t (A_t + \mu \cot \! \beta) \\
- m_t (A_t + \mu \cot \! \beta ) & m^2_{\tilde{t}_R} + m_t^2 + 0.16 D_Z
\end{array} \right) $}; \label{e4a} \\
{\cal M}^2_{\tilde b} &= \mbox{$ \left( \begin{array}{cc}
m^2_{\tilde{t}_L} + m_b^2 - 0.42 D_Z & - m_b (A_b + \mu \tanb) \\
- m_b (A_b + \mu \tanb) & m^2_{\tilde{b}_R} + m_b^2 - 0.08 D_Z
\end{array} \right) $}, \label{e4b} \eeq \een
where $D_Z = M_Z^2 \cos \! 2 \beta$, $\tan \beta$ being the ratio of the
vacuum expectation values of the two neutral Higgs fields of the MSSM
\cite{mssm}. $m_{\tilde{t}_L,\tilde{t}_R,\tilde{b}_R}$ are soft breaking
masses, $A_{b,t}$ are parameters describing the strength of nonsupersymmetric
trilinear scalar interactions, and $\mu$ is the supersymmetric Higgs(ino)
mass, which also enters trilinear scalar vertices. Notice that the
off--diagonal elements of these squark mass matrices are proportional to the
quark mass. In the case of the supersymmetric partners of the light quarks
mixing between the current eigenstates can therefore be neglected. However,
mixing between $\tilde{t}$ squarks can be sizable and allows one of the mass
eigenstates to be much lighter than the top quark. Sbottom mixing can also be
significant if $\tanb \gg 1$; even in supergravity models with radiative
symmetry breaking \tanb\ can be as large as $m_t/m_b$ \cite{dn1}. \s

In the presence of squark mixing, the squark-quark-gluino interaction
Lagrangian is given by
\begin{eqnarray} \label{e7}
{\cal L}_{\tilde{g}\tilde{q}q} = -i\sqrt{2}g_s T^a \overline{q} \left[ \left(
\cos\heta \tilde{q_1} - \sin \heta \tilde{q_2} \right) \frac{1+\gamma_5}{2} -
\left( \sin \heta \tilde{q_1} + \cos  \heta \tilde{q_2}
\right)\frac{1-\gamma_5}
{2} \right] \tilde{g}_a + {\rm h.c.},
\end{eqnarray}
where $g_s$ is the strong coupling constant and $T^a$ are SU(3)$_C$
generators. Note that in eq.~(\ref{e7}) we have assumed $M_{\tilde{g}}>0$. \s

\subsection*{2a. Higgs boson decays into light gluinos}

Summing over colors and taking into account the fact that there are two
identical particles in the final state, the partial decay widths of the
CP--even Higgs bosons $H_1, H_2 \equiv h$ and the CP--odd boson $H_3 \equiv P$
into a pair of gluinos in the limit $\mg \rightarrow 0$ are given by
\begin{eqnarray}
\Gamma (H_k \rightarrow \sgg) = \frac{1}{2} \alpha M_{H_k}\left(\frac{\alpha_S}
{\pi} \right)^2 \left( \sum_{q=t,b} A_k^q \right)^2.
\end{eqnarray}
The amplitudes $A_k^q$ can be written as
\begin{eqnarray} \label{e2.1}
A_{1,2}^q &=& \frac{1}{2} (s_{1,2})^q \sin 2\theta_q \left[ (m_q^2 +
m_{\tilde{q}_2}^2) C_0(m_q,m_q,m_{\tilde{q}_2}) - (m_q^2+m_{\tilde{q}_1}^2)
C_0(m_q,m_q,m_{\tilde{q}_1}) \right] - \frac{1}{2} m_q \sin 2 \theta_q \non \\
& &  \times \left[\left( \tilde{s}^{11}_{1,2}\right)^q C_0(m_{\tilde{q}_1},
m_{\tilde{q}_1},m_q)
-\left( \tilde{s}^{22}_{1,2} \right)^q C_0(m_{\tilde{q}_2},m_{\tilde{q}_2},m_q)
+2 \left(
\tilde{s}^{12}_{1,2} \right)^q \cot 2\theta_q C_0(m_{\tilde{q}_1},
m_{\tilde{q}_2},m_q) \right]; \non
\end{eqnarray}
\begin{eqnarray}
A_3^q &=& \frac{1}{2} (s_{3})^q \sin 2\theta_q \left[
(m_q^2 - m_{\tilde{q}_2}^2) C_0(m_q,m_q,m_{\tilde{q}_2}) -
(m_q^2 - m_{\tilde{q}_1}^2)C_0(m_q,m_q,m_{\tilde{q}_1})\right]
\non \\
& &  + \left( \tilde{s}^{12}_3 \right)^q m_q C_0(m_{\tilde{q}_1},
m_{\tilde{q}_2},m_q),
\end{eqnarray}
where the scalar function $C_0(m_1,m_2,m_3)$ is given in integral form by
\begin{eqnarray}
C_0(m_1,m_2,m_3) = - \int_0^1 dy \int_0^y dx \left[ M_{H_k}^2 x(x-y) +
(m_2^2-m_3^2)y + (m_1^2-m_2^2)x+ m_3^2-i \epsilon \right]^{-1}.
\end{eqnarray}
A complete expression after integration over the Feynman variables is given in
the Appendix. In the limit $ M_{H_k}^2 \ll m_t^2$ this function simplifies to
\begin{eqnarray}
C_0(m_1,m_2,m_3) = \frac{1}{m_2^2-m_1^2} \left[ \frac{m_1^2\log m_1^2
-m_3^2\log m_3^2} {m_1^2-m_3^2} -\frac{m_2^2\log m_2^2
-m_3^2\log m_3^2} {m_2^2-m_3^2} \right],
\end{eqnarray}
and for $m_1=m_2$
\begin{eqnarray} \label{e2.2}
C_0(m,m,m_3) = \frac{1}{m^2_3-m^2}+\frac{m_3^2}{(m^2_3-m^2)^2}
\log\frac{m^2}{m_3^2}
\end{eqnarray}
The $H_k q q$ couplings $(s_k)^q$ are given by \cite{mssm}
($s_W^2=1-c_W^2=\sin^2\theta_W$)
\begin{eqnarray}
(s_k)^q= \frac{m_q}{2s_W M_W} r_k^q,
\end{eqnarray}
with
\begin{eqnarray}
r_1^t = \frac{\sin \alpha}{\sin \beta} \ \ , \
r_2^t = \ \frac{\cos \alpha}{\sin \beta} \ \ , \
r_3^t = \frac{1}{\tanb}; \non \\
r_1^b = \frac{\cos \alpha}{\cos \beta} \ \ , \
r_2^b = -\frac{\sin \alpha}{\cos \beta} \ \ , \
r_3^b = \tanb.
\end{eqnarray}
The $H_k \tilde{q}_i \tilde{q}_j$ couplings $(\tilde{s}^{ij}_k)^q$ are
\cite{dn7}
\begin{eqnarray}
(\tilde{s}_1^{11})^q &=& \frac{M_Z \cos (\alpha+\beta)}{s_Wc_W} \left( I_{3q}^L
\cos^2\theta_q- e_qs_W^2 \cos 2\theta_q \right)  + \frac{m_q^2 r_1^q}{s_WM_W}
-\frac{m_q \sin2\theta}{2s_W M_W} (r_1^q A_q + r_2^q \mu);\non \\
(\tilde{s}_1^{22})^q &=& \frac{M_Z \cos(\alpha+\beta) }{s_Wc_W} \left( I_{3q}^L
\sin^2\theta_q+ e_qs_W^2 \cos 2\theta_q \right) + \frac{m_q^2r_1^q}{s_WM_W}
+\frac{m_q \sin2\theta }{2s_W M_W} (r_1^q A_q + r_2^q \mu); \non \\
(\tilde{s}_1^{12})^q &=& \frac{M_Z\cos(\alpha+\beta)
}{s_Wc_W} \sin 2\theta_q \left( e_q s_W^2 -\frac{1}{2}I_{3q}^L \right)
-\frac{m_q}{2s_W M_W} (r_1^q A_q + r_2^q \mu) \cos 2\theta_q;
\end{eqnarray}
\begin{eqnarray}
(\tilde{s}_2^{ij})^q = (\tilde{s}_1^{ij})^q \ [\sin \alpha \rightarrow + \cos
\alpha \ , \cos \alpha \rightarrow - \sin \alpha ];
\end{eqnarray}
\begin{eqnarray} \label{e13}
(\tilde{s}_3^{11})^q =  (\tilde{s}_3^{22})^q =0 \ \ \ , \
(\tilde{s}_3^{12})^q = \frac{m_q}{2s_W M_W}
(\mu - r_3^q A_q).
\end{eqnarray}

\nn A few remarks are in order: \s

$(i)$ As already mentioned, eqs.(\ref{e2.1})--(\ref{e2.2}) are valid only if
the gluinos are nearly massless; this is an excellent approximation for the
case of interest, $\mg \leq 5$ GeV $\ll m_t, m_{H_k}, m_{\tilde{q}}$. Complete
expressions for a finite gluino mass are given in the Appendix. Note that the
contribution of diagram Fig.~1b is ultraviolet finite but the contribution of
diagram Fig.~1a is finite only once the contributions of both squarks of a
given flavor have been added. \s

$(ii)$ The main conventional decay mode of Higgs particles with masses
below 130 GeV is the decay into $b\bar{b}$ pairs; in the limit $M_{H_k}\gg
m_b$, the corresponding decay width is given by
\begin{eqnarray}
\Gamma (H_k \rightarrow b\bar{b}) = \frac{3}{2} \alpha M_{H_k} \left(
\frac{m_b}{2s_WM_W} \right)^2 (r_k^b)^2.
\end{eqnarray}
Note that one has to include the QCD corrections to this decay width. The bulk
of these corrections can be absorbed \cite{corr} into running quark masses
evaluated at the scale $\mu=M_{H_k}$. For Higgs masses arround 100 GeV, the
$b$--quark mass $m_b(m_b)=4.5$ GeV will drop to the effective value $m_b
(M_{H_k}) \simeq 3.2$ GeV; this results in a decrease of the decay width by
approximately a factor of two.\footnote{In contrast to the running of \as,
light gluinos have very little effect on the running of $m_b$.} \s

$(iii)$ In the limit where either $A_q$ and $\mu$ or $m_q$ are set to zero,
there is no mixing between left-- and right--handed squarks. In this case, the
amplitudes $A_k^q$ are $\propto \mg$ and hence very small; see also
ref.~\cite{ng}. Therefore only the contribution of the top quark and its SUSY
partners have to be taken into account; (s)bottom loop contributions are
sizable only if $\tanb \gg 1$. \s

\subsection*{2b. The decay $Z \rightarrow \sgg $}

The partial decay width of the $Z$ boson into a pair of massless gluinos is
given by
\begin{eqnarray}
\Gamma (Z \rightarrow \sgg) = \frac{\alpha M_Z}{48 c_W^2 s_W^2} \left(
\frac{\alpha_S} {\pi} \right)^2 \left[ \sum_q \left(
\sum_{i=1}^2  B_i^q + \sum_{i,j=1}^2  \tilde{B}_{ij}^q \right) \right]^2,
\end{eqnarray}
where [here $s=M_Z^2$]
\begin{eqnarray}
B_i^q &=& (v_i^2+a_i^2)a_q \left[ 2C_2(s,m_q,m_q,\msi)-B_0(s,m_q,m_q)
-(m_q^2+\msi^2) C_0(s,m_q,m_q,\msi) \right] \non \\
&& + 2 (v_ia_i)v_q \left[ 2C_2(s,m_q,m_q,\msi)-B_0(s,m_q,m_q)+
(m_q^2-\msi^2) C_0(s,m_q,m_q,\msi) \right]; \non \\
\tilde{B}_{ij}^q &=& - 2 a_{ij} (v_ia_j+ v_ja_i) C_2(s,\msi,\msj,m_q),
\end{eqnarray}
\nn with
\begin{eqnarray}
a_{11}= 2(2I_q^{3L} \cos^2 \heta -2s_W^2 e_q) \ , \
a_{22}= 2(2I_q^{3L} \sin^2 \heta -2s_W^2 e_q) \ , \
a_{12}= -2 I_q^{3L} \sin 2\heta,
\end{eqnarray}
\begin{eqnarray}
v_q= 2I_q^{3L} -4 s_W^2 e_q \ \ & , & \ \  a_q= 2I_q^{3L},
\end{eqnarray}
and
\begin{eqnarray}
v_1= \frac{1}{2}(\cos \theta_q - \sin \theta_q ) = a_2 \ \ \ , \ \
a_1= \frac{1}{2}(\cos \theta_q + \sin \theta_q ) = -v_2.
\end{eqnarray}

Alternatively, the result can be written as
\begin{eqnarray}
\Gamma (Z \rightarrow \sgg) = \frac{\alpha M_Z}{48 c_W^2 s_W^2} \left(
\frac{\alpha_S} {\pi} \right)^2 \left( \sum_q B^q \right)^2,
\end{eqnarray}
with
\begin{eqnarray}
B^q &=& \frac{1}{2}(a_q+v_q \cos2\heta) \left[2C_2(s,m_q,m_q,\tilde{m}_{q_1})
-B_0(s,m_q,m_q) +(m_q^2-\tilde{m}^2_{q_1}) C_0(s,m_q,m_q,\tilde{m}_{q_1})
\right] \non \\
&+& \frac{1}{2}(a_q-v_q \cos2\heta) \left[2C_2(s,m_q,m_q,\tilde{m}_{q_2})
-B_0(s,m_q,m_q) +(m_q^2-\tilde{m}^2_{q_2}) C_0(s,m_q,m_q,\tilde{m}_{q_2})
\right] \non \\
&-& a_q m_q^2 \left[ C_0(s,m_q,m_q,\tilde{m}_{q_1})
+C_0(s,m_q,m_q,\tilde{m}_{q_2}) \right]  \\
&-& \cos 2\heta \left[a_{11} C_2(s,\tilde{m}_{q_1}, \tilde{m}_{q_1},m_q)
-a_{22} C_2(s,\tilde{m}_{q_2}, \tilde{m}_{q_2},m_q) \right] \non \\
&+& 2 \sin 2\heta a_{12} C_2(s,\tilde{m}_{q_1}, \tilde{m}_{q_2},m_q). \non
\end{eqnarray}

In terms of the scalar two and three point functions $B_0$ and
$C_0$ given in the Appendix, the function $C_2$ is defined as
\begin{eqnarray}
C_2(s,m_1,m_2,m_3) = \frac{1}{2}m_3^2 C_0(s,m_1,m_2,m_3) +\frac{1}{4}
+\frac{1}{4}
B_0(s,m_1,m_2)-\frac{1}{8s}(m_1^2+m_2^2-2m_3^2) \hspace*{0.7cm} \non \\
\ \ \ \times \left[ 2B_0(s,m_1,m_2)- B_0 (0,m_3,m_1) -B_0 (0,m_3,m_2)
+ (2m_3^2 -m_1^2-m_2^2) C_0(s,m_1,m_2,m_3) \right] \non \\
+\frac{1}{8s}(m_1^2-m_2^2) \left[ B_0 (0,m_3,m_1) -B_0 (0,m_3,m_2)
+ (m_2^2-m_1^2) C_0(s,m_1,m_2,m_3) \right] \hspace*{0.8cm}
\end{eqnarray}

Note that $B_i^q$ and $\tilde{B}_{ij}^q$ are ultraviolet divergent [the term
proportional to the vectorial coupling $v_q$ in $B_i^q$ is finite when one
sums over the contributions of the two squarks of a given flavor]. It is only
when one sums over a complete isodoublet that the amplitudes are finite. The
contributions from (s)bottom loops therefore always have to be included here.
On the other hand, if both squarks and quarks of a given generation are
degenerate in mass, this generation will not contribute to $B_q$; see also
refs.~\cite{7,norway}.

\subsection*{3. Numerical Results}

We are now in a position to present numerical results for branching ratios of
$Z$ and Higgs bosons into pairs of light gluinos. As stated in the
Introduction, we will assume minimal (s)particle content, i.e. two Higgs
doublets as well as the usual gauge and matter superfields. Moreover, unless
stated otherwise we will assume that explicitly SUSY breaking contributions to
the squark mass matrices are identical for all flavors. In terms of the
parameters introduced in eqs.(\ref{e4}), this implies:
\ben \label{e3.1} \beq
m_{\tilde{t}_L} &= m_{\tilde{t}_R} = m_{\tilde{b}_R} \equiv m_{\tilde{q}};
\label{e3.1a} \\
A_t &= A_b \equiv A \cdot m_{\tilde{q}}. \label{e.31b} \eeq \een
When computing Higgs branching fractions we have to specify one more
parameter, besides those appearing in eqs.(\ref{e4}). We choose this to be the
mass $m_P$ of the pseudoscalar Higgs boson. We include full 1--loop
corrections \cite{l1} from the (s)top and (s)bottom sectors to the mass and
mixing angle of the scalar Higgs bosons, including non--logarithmic terms
\cite{l2}. \s

The values of these parameters are constrained by unsuccessful searches for
squarks and Higgs bosons. However, many bounds that have been derived under
the assumption that gluinos are heavy are no longer valid if the gluino mass
is just a few GeV. In particular, heavy squarks would almost always decay into
the corresponding quark plus a gluino. Since a light gluino will lose a
substantial fraction of its energy in radiation prior to its decay \cite{3},
the missing transverse momentum in such events might be too small for them to
pass cuts designed for conventional squark searches at hadron colliders
\cite{pp}. On the other hand, LEP experiments allow to place bounds on squark
masses simply from the measurement of the total and hadronic widths of the $Z$
boson, independent of how the squarks decay. We therefore require all squark
mass eigenstates to be heavier than 45 GeV.\footnote{In practice we only
include contributions from third generation (s)quarks; large masses for
squarks of the first and second generation would not affect our results. This
makes it even less likely that the parameter space relevant for us is
constrained significantly by sparticle searches at $p \overline{p}$ colliders.}
\s

Turning to bounds on the Higgs sector, we note that searches for $Z
\rightarrow Z^* h$ decays (the Higgs bremsstrahlung process) are little
affected by Higgs decay modes as long as it decays hadronically. We have
therefore excluded combinations of parameters that violate the
$m_h$--dependent bound on the $ZZh$ coupling derived by the ALEPH
collaboration \cite{aleph}.\footnote{The OPAL collaboration has recently
published improved Higgs search limits \cite{opal}. Unfortunately the
information given in their paper does not allow to extract an upper bound on
the $ZZh$ coupling.} On the other hand, searches for $Z \rightarrow h P$
decays do make use of the assumption that Higgs bosons accessible at LEP
decay predominantly into pairs of $b$ quarks or $\tau$ leptons; some sort of
heavy fermion tagging is necessary in order to suppress the QCD 4--jet
background. Thus bounds from searches for associate $hP$ production may not
apply if these Higgs bosons have a significant or even dominant branching
ratio into light gluinos. We have therefore not included these bounds in our
analysis. \s

In figs.~2a,b we show results for $Br(h \rightarrow \glgl)$ for $m_{\tilde{q}}
=400$ GeV, $\mu = 200$ GeV and several combinations of $m_P$ and \tanb. The
curves have been obtained by varying the $A$ parameter in the region $A<0$
from its minimal allowed value, defined by $m_{\tilde{t}_1} = 45$ GeV, to the
point where $m_{\tilde{t}_1}$ is maximized ($A_t = -\mu \cot \! \beta$). The
results are presented as a function of the light stop mass $m_{\tilde{t}_1}$,
since it is more directly accessible experimentally than the $A$ parameter.
\s

We find that for $m_P^2 \gg m_Z^2$ the dependence of the branching ratio on
either $m_P$ or \tanb\ is quite weak. In this limit the light Higgs scalar $h$
couples to quarks, leptons and massive gauge bosons with the same strength as
the single physical Higgs boson of the SM \cite{mssm,rev}. If in addition
$m^2_{\tilde{t}_1} \ll m^2_{\tilde{q}}$, the $h \tilde{t}_1 \tilde{t}_1$
coupling $\left( s_2^{11} \right)^q$ is \cite{dn7} proportional to the
off--diagonal element of the stop mass matrix (\ref{e4a}); this implies that
$\left( s_2^{11} \right)^q \propto \left(m^2_{\tilde{t}_2} -
m^2_{\tilde{t}_1} \right)/m_W \propto m^2_{\tilde{q}}/m_W$ in this limit. For
fixed $m_{\tilde{t}_1}$, the square of this coupling, and hence $\Gamma(h
\rightarrow \glgl)$, therefore grows like the fourth power of \msq. This
explains why decays into light gluinos completely dominate all other decay
modes for small $m_{\tilde{t}_1}$ and large $m_P$, as shown in figs.~2. Even
for the rather modest value of \msq\ chosen here this new deay mode can
suppress the branching ratios into $b \bar{b}$ or $\tau^+ \tau^-$ pairs by
more than a factor of 1,000.
\s

The situation is more complicated for smaller values of $m_P$. If $m_P$ is
of order $m_Z$ or less, the scalar mixing angle changes such that $\Gamma
(h \rightarrow b \bar{b}) \propto \tan^2 \beta$ for large \tanb, while the
coupling of $h$ to top quarks becomes $\propto \cot \! \beta$. Both effects
lead to a rapid decrease of $Br(h \rightarrow \glgl)$ if $m_P$ is not large.
Even for \tanb\ as small as 2, fig.~2b, for small $m_P$ the gluino mode is
important only if $m_{\tilde{t}_1}$ is very small, although it might still
remain measurable over a wider region of parameter space. Notice also that
the branching ratio now depends on $A_t$ and $\mu$ separately, not only on
the combination $A_t + \mu \cot \! \beta$ that appears in the stop mass
matrix (dotted and short dashed curves).
\s

Fig.~3 shows results for the decay of the pseudoscalar Higgs boson into light
gluinos. The couplings of $P$ are uniquely determined by \tanb, independent
of $m_P$; our result is therefore almost independent of $m_P$ as long as
it lies in the region $2 m_b \leq m_P \leq 2 m_t$.\footnote{We include $P
\rightarrow Z h$ decays where kinematically allowed when computing the total
decay width of $P$, but this contribution is usually very small.} In fig.~3
we took $m_P=50$ GeV, and chose the same values of \msq\ and $\mu$ as in
fig.~2.
\s

We see from fig.~3 that the maximal contribution of the \glgl\ final state to
the total decay width of $P$ is clearly much smaller than in case of the
scalar $h$. The reason is that there is no $P \tilde{t}_1 \tilde{t}_1$
coupling \cite{mssm}. The $P \tilde{t_1} \tilde{t_2}$ coupling (\ref{e13})
can still become quite large if $m_{\tilde{t}_1} \ll \msq$; however, this
coupling only contributes through a diagram containing one {\em heavy} stop
$\tilde{t}_2$ in the loop, giving a factor $1/m^2_{\tilde{t}_2}$ from the
loop integration. As a result, for fixed $m_{\tilde{t}_1}$ the $P \rightarrow
\glgl$ partial width becomes independent of \msq\ once $m^2_{\tilde{q}} \gg
m^2_t$, as compared to a growth $\propto m^4_{\tilde{q}}$ in case of the
scalar Higgs boson $h$. Nevertheless, for small \tanb\ the \glgl\ mode can
still be quite important even for the pseudoscalar, suppressing the branching
ratio into $b$ and $\tau$ pairs by a factor of about 3. For larger \tanb\ the
width for these standard modes grows $\propto \tan^2 \beta$, while the
$Ptt$ and $P \tilde{t}_1 \tilde{t}_1$ couplings decrease; the \glgl\ mode
therefore becomes negligible for $\tanb > 5$ or so.
\s

The same parameters that determine the partial widths for Higgs boson decays
into light gluinos also determine $\Gamma(Z \rightarrow \glgl)$; one might
therefore hope to place bounds on the former by constraining the latter.
However, fig.~4 shows that $\Gamma (Z \rightarrow \glgl)$ depends only very
weakly on details of $L-R$ squark mixing, described by the
parameters $A, \ \mu$ and \tanb. Recall that we assume squarks of the first
two generations to be degenerate in mass; these generations do therefore not
contribute to the $Z \rightarrow \glgl$ amplitude. Mixing, as well as
splitting of mass eigenstates, can be sizable for stop squarks; we have seen
above that the partial widths for Higgs decays into light gluinos strongly
depend on these details of the stop sector. In contrast, the $Z \tilde{t}_i
\tilde{t}_j$ couplings are pure gauge couplings, which are usually modified
only weakly by squark mixing. Moreover, reducing $m_{\tilde{t}_1}$ below $m_t$
has little effect on the loop integrals. It is also important to keep in mind
that we get a meaningful result only after summing over all four squark mass
eigenstates of a given generation; this summation further reduces the
dependence on the details of the squark mass matrices. Even increasing
$m^2_{\tilde{t}_R}$ and $m^2_{\tilde{b}_R}$ to $2 \cdot m^2_{\tilde{t}_L}$
(dotted curve) reduces $\Gamma(Z \rightarrow \glgl)$ by at most 20\%, since
the couplings of right--handed squarks to the $Z$ are rather weak anyway.
\s

We see that within our assumptions $Br(Z \rightarrow \glgl) \leq 7 \cdot
10^{-4}$. Somewhat larger branching ratios are possible \cite{norway} if we
allow mass splitting between squarks of the first two generations, but this
splitting is tightly constrained by bounds on flavor--changing neutral
currents, most notably $K^0 \overline{K^0}$ mixing \cite{fcnc}. The
contribution of light gluinos to the total or hadronic decay width of the $Z$
boson is thus always well below the present experimental error.\footnote{
Light gluinos will also contribute to the hadronic and total $Z$ boson width
through loop diagrams; however these contributions are generally small
\cite{DDK}.} One will
therefore have to devise cuts that allow to distinguish gluino pairs from
pairs of light quarks. One obvious example is to seach for events with missing
energy. However, this requires knowledge of the gluino fragmentation
function which is not very well understood if gluinos are light \cite{3}.
Moreover, there are substantial backgrounds, both instrumental (from the
mismeasurement of jet energies) and irreducible (from the production and
subsequent semi--leptonic decay of heavy quarks).
\s

A more promising strategy might therefore be \cite{cuypers} to search for
$\tilde{g}$ pairs by looking for their decay vertices. Farrar recently
concluded \cite{farrar} that $\mg \geq 3$ GeV, mostly due to constraints from
searches for radiative $\Upsilon$ decays into $\glgl$ bound states; the decay
vertices of gluinos of this mass should be readily detectable, unless the LSP
mass is very close to the gluino mass. If the LSP is an (almost) massless
photino, the gluino lifetime is approximately given by \cite{farrar}
\be \label{e3.2}
\tau_{\tilde{g}} \simeq 5 \cdot 10^{-18} \ {\rm sec} \cdot \frac{3 \ {\rm
GeV}} {\mg} \left( \frac {\msq}{\mg} \right)^4. \ee
This gives a mean flight path of approximately 0.1 mm (2 cm) for $\msq=50$
(100) GeV and $\mg=3$ GeV. A detailed Monte Carlo study will be necessary to
determine whether detection of light gluinos at LEP via these or other
\cite{lepgl,cuypers} methods is feasible.
\s

Finally, in figs.~5 a--c we show scatter plots in the planes spanned by any
two of the three branching ratios discussed above. These plots have been
obtained by randomly choosing sets of input parameters within the limits
50 GeV $\leq \msq \leq$ 500 GeV, 50 GeV $\leq \mu \leq$ 1 TeV, 25 GeV $\leq
m_P \leq$ 500 GeV, $1.2 \leq \tanb \leq 25$ and $-4 \leq A \leq 4$, excluding
combinations of parameters that lead to too light a squark eigenstate or too
large a $ZZh$ coupling. Fig.~5a shows a weak positive correlation between the
$h$ and $P$ branching ratios: Most points with $Br(\Pgl) > 10\%$ also have
$Br(\hgl) > 1 \%$ at least; there are some counter--examples, however. Also,
the number of points with $Br(\hgl) > 10 \%$ is clearly much larger than that
with $Br(\Pgl) > 10 \%$; this is not surprising, given the very different
dependence of the two decay amplitudes on \tanb\ and on \msq, as discussed
above. \s

Fig.~5b shows that the $h$ and $Z$ branching ratios into light gluinos are
anti--correlated. The reason is that, as explained above, a large $Br(\hgl)$
can most easily be obtained if $m^2_{\tilde{q}} \gg m^2_{\tilde{t}_1}$ is
sizable, which suppresses $\Gamma(\Zgl)$. Fig.~5c shows no significant
correlation between the $P$ and $Z$ branching ratios into gluinos. Taken
together, figs.~5 a--c demonstrate that the measurement of any one of the
three branching ratios discussed here would not allow one to predict, or even
significantly constrain, the other two.

\subsection*{4. Summary and Conclusions}

In this paper we computed the decay widths of Higgs and $Z$ bosons into pairs
of light gluinos. Although this decay only occurs at the 1--loop level, it can
dominate the total decay width of the light scalar Higgs boson by a large
factor if the off--diagonal entry of the stop mass matrix is large, in which
case the lighter stop eigenstate is expected to be (much) lighter than the
other squarks. The contribution of the \glgl\ final state to the total decay
width of the pseudoscalar Higgs boson can also be quite sizable, although not
as large as for the sclar Higgs boson. In contrast, the branching ratio for
\Zgl\ decays is always below one permille; a dedicated search will be
necessary to explore this possibility experimentally. Unfortunately placing
bounds on, or a measurement of, $Br(\Zgl)$ will not teach us much about Higgs
branching ratios into light gluinos. \s

What are the consequences of large $Br(h,P \rightarrow \glgl)$? As discussed
in sec.~3, searches for $Z \rightarrow Z^* h$ decays at LEP1, or $e^+e^-
\rightarrow Zh$ production at LEP2, are probably not much affected, since here
it is usually not necessary to tag the Higgs decay products in order to
isolate a detectable signal. Some heavy fermion tagging {\em is} necessary in
searches for associate $hP$ production, however. Notice that at present only
the combined limits on $Z^*h$ and $hP$ production allow one to place lower
bounds on $m_h$ and $m_P$ \cite{aleph,opal}. We are therefore forced to
conclude that the present bounds on the masses of the Higgs bosons of the MSSM
may well be invalid if there are indeed gluinos with mass of only a few GeV. \s

On the other hand, in the long run the existence of light gluinos need not
hamper Higgs searches at $e^+ e^-$ colliders. While gluino pairs do look
different from $b$ and especially $\tau$ pairs, they also ought to differ
sufficiently from standard light quark or gluon jets to allow the suppression
of QCD backgrounds. The impact of light gluinos on Higgs searches at hadron
colliders might be less benign, however. The huge background from pure QCD
processes means that gluino pairs cannot be used as a Higgs signal here. We
have seen that in the presence of light gluinos the branching ratio of scalar
Higgs bosons into $b$ or $\tau$ pairs might be reduced by a factor of 1,000 or
more. The same suppression factor applies for decays into (virtual) gauge
bosons. The $h \rightarrow ZZ^* \rightarrow 4$ leptons signal would then
be undetectable; however, within the MSSM it is at best marginal anyway
\cite{hpp}. \s

Potentially most unfortunate for the prospects of MSSM Higgs searches at
hadron colliders would be a reduction of the $h \rightarrow \gamma \gamma$
signal, which is usually regarded to be the most promising way to search for
$h$ \cite{hpp}. However, the same $h \tilde{t}_1 \tilde{t}_1$ coupling that
can give rise to a large partial width for the \glgl\ final state also
contributes to the $h \rightarrow \gamma \gamma$ and $h \rightarrow gg$
partial widths, via $\tilde{t}_1$ loops. A proper investigation of the
possible impact of the existence of light gluinos on Higgs boson searches at
hadron colliders therefore necessitates an analysis of all SUSY loop
contributions to Higgs production and decay, which is beyond the scope of this
paper. We nevertheless hope that our results will give additional urgency to
the effort to either detect or definetely exclude the existence of gluinos
with mass of a few GeV. \s

\vspace*{1cm}
\noindent
{\bf Acknowledgements}
We thank D. Summers, as well as the members of the OPAL group of the
Universit\'e de Montr\'eal for discussions. M.D. thanks the Groupe de Physique
des Particules of the Universit\'e de Montreal for their hospitality while
this project was initiated. His work was supported in part by the U.S.
Department of Energy under contract No. DE-AC02-76ER00881, by the Wisconsin
Research Committee with funds granted by the Wisconsin Alumni Research
Foundation, as well as by a grant from the Deutsche Forschungsgemeinschaft
under the Heisenberg program. A.D. was supported by an NSERC fellowship.

\newpage

\renewcommand{\theequation}{A.\arabic{equation}}
\setcounter{equation}{0}
\subsection*{Appendix: Complete result}

Taking into account the mass of the gluino, the partial decay width of the
CP--even and CP--odd Higgs bosons into gluino pairs is given by
\begin{eqnarray}
\Gamma (H_k \rightarrow \sgg) = \frac{1}{2} \alpha M_{H_k}\left(\frac{\alpha_S}
{\pi} \right)^2 \beta_{\tilde{g}}^p \left[ \sum_q \left( \sum_{i=1}^2 A^{i}_k
+ \sum_{i,j=1}^2  \tilde{A}^{ij}_k \right) \right]^2
\end{eqnarray}
where $\beta_{\tilde{g}}=1-4M_{\tilde{g}}^2/M_{H_k}^2$ is the velocity of the
final gluinos with $p=1$ for $k=1,2$ and $p=3$ for $k=3$. The amplitudes
$A^{i}_k$ come from the diagrams Fig.~1a and $\tilde{A}^{ij}_k$ come from
the diagrams Fig.~1b; they are given by

\begin{eqnarray}
A^i_{1,2} &=& (s_{1,2})^q (v_i^2-a_i^2) \left[ B_0(s,m_q,m_q) + (\mg^2+m_q^2+
\msi^2) C_0(s,m_q,m_q,\msi) + 4\mg^2 \right.  \non \\
& & \left. C_+(s,m_q,m_q,\msi) \right] + 2 m_q \mg (v_i^2+a_i^2)
\left[ C_0(s,m_q,m_q,\msi)+2 C_+(s,m_q,m_q,\msi) \right] \non \\
A^i_{3} &=& (s_{3})^q (v_i^2-a_i^2) \left[ - B_0(s,m_q,m_q) + (\mg^2+m_q^2 -
\msi^2) C_0(s,m_q,m_q,\msi) \right] \\
& & + 2 m_q \mg (v_i^2+a_i^2) C_0(s,m_q,m_q,\msi)
\non \\
\tilde{A}^{ij}_{1,2} &=& \left( \tilde{s}_{1,2}^{ij} \right)^q \left[ m_q(
v_iv_j-a_ia_j) C_0(s,\msi,\msj,m_q) -2\mg (v_i v_j+a_ia_j)
C_+(s,\msi,\msj,m_q) \right]
\non \\
\tilde{A}^{ij}_{3} &=& \left( \tilde{s}_{3}^{ij} \right)^q \left[ m_q (v_ia_j-
v_j a_i) C_0(s,\msi,\msj,m_q) - 2\mg (v_i a_j+a_iv_j) C_-(s,\msi,\msj,m_q)
\right] \non
\end{eqnarray}
with the couplings $(s_k)^q$ and $(\tilde{s}_k^{ij})^q$ given in eqs.~(6--10)
and
\begin{eqnarray}
v_1= \frac{1}{2}(\cos \theta_q - \sin \theta_q ) = a_2 \ \ \ , \ \
a_1= \frac{1}{2}(\cos \theta_q + \sin \theta_q ) = -v_2
\end{eqnarray}

\nn The functions $C_+(s,m_1,m_2,m_3)$ is defined as
\begin{eqnarray}
C_+(s,m_1,m_2,m_3) &=& \frac{1}{2 s \beta_{\tilde{g}}^2} \left[ 2B_0(s,m_1,
m_2)- B_0 (\mg^2,m_3,m_1) -B_0 (\mg^2,m_3,m_2) \right. \non \\
&& \left. \hspace*{1cm} + (2 \mg^2+ 2m_3^2 -m_1^2-m_2^2)C_0(s,m_1,m_2,m_3)
\right] \\
C_-(s,m_1,m_2,m_3) &=& \frac{1}{2s} \left[
B_0(\mg^2,m_3,m_2)-B_0(\mg^2,m_3,m_1)
+ (m_1^2-m_2)^2 C_0(s,m_1,m_2,m_3) \right] \non
\end{eqnarray}
with the scalar two and three point functions, $B_0$ and $C_0$ defined as
\begin{eqnarray} \label{a1}
B_0(s,m_1,m_2) &=& \frac{(2\pi\mu)^{n-4}}{i\pi^2} \ \int \frac{d^nk}{
(k^2-m_1^2 + i\epsilon) [(k-q)^2-m_2^2+i \epsilon] }, \\
C_0(s, m_1, m_2, m_3) &=& \frac{(2\pi\mu)^{n-4}}{i\pi^2} \int \frac{d^nk}{
[(k-p_1)^2-m_1^2 +i\epsilon][(k-p_2)^2-m_2^2 +i\epsilon](k^2-m_3^2+i\epsilon)}.
\non
\end{eqnarray}

\nn where $n$ is the space--time dimension and $\mu$ the renormalisation scale.
After integration over the internal momentum $k$, the function $B_0$
is given by [$\gamma_E$ is Euler's constant]
\begin{eqnarray} \label{a3}
B_0(s,m_1,m_2) &=& \frac{1}{\epsilon} -\gamma_E + + \log \frac{4\pi\mu^2 }
{m_1 m_2} +2 +\frac{m_1^2-m_2^2}
{2s} \log \frac{m_2^2}{m_1^2} +\frac{x_+ -x_-}{4s} \log \frac{x_-}{x_+} \non
\end{eqnarray}
\nn with
\begin{eqnarray} \label{a4}
x_\pm =s- m_1^2 -m_2^2 \pm \sqrt{ s^2-2s (m_1^2+m_2^2)+(m_1^2 - m_2^2 )^2}
\end{eqnarray}

\nn The three point scalar function $C_0$ for $p_1^2 = p_2^2 =
m_{\tilde{g}}^2$ and $s = \left( p_1 + p_2 \right)^2$ is given in integral
form by
\begin{eqnarray}
C_0(s,\mg,m_1,m_2,m_3) = - \int_0^1 dy \int_0^y dx \left[ ay^2+bx^2+cxy
+dy+ex+f \right]^{-1},
\end{eqnarray}
where
\begin{eqnarray} \label{a6}
a=\mg^2 \ , \ \ b=s \ , \ \ c=-s \ , \ \ d = m_2^2 -m_3^2 - \mg^2 \ ,
\ \ e = m_1^2 -m_2^2 \ , \ \ f=m_3^2 -i \epsilon. \hspace*{3mm}
\end{eqnarray}

\nn $C_0$ can be expressed in terms of a sum of Spence functions
Li$_2(x)=-\int_0^1 dt \log(1-xt)/t$:
\begin{eqnarray} \label{a7}
C_0(s,\mg,m_1,m_2,m_3)= - \frac{1}{s \beta_{\tilde{g}}} \ \sum_{i=1}^{3}
\sum_{j=+,-}
\ (-1)^i \left[ {\rm Li_2} \left( \frac{x_i} {x_i-y_{ij} } \right) -
{\rm Li_2} \left( \frac{x_i-1}{x_i-y_{ij}} \right) \right],
\end{eqnarray}
where we have defined
\begin{eqnarray} \label{a8}
x_1=\frac{2d+e(1-\beta_q)}{2s\beta_q}+\frac{1}{2}(1-\beta_q) \ \ ,
\hspace*{1cm} & & y_{1\pm} =\frac{-c-e \pm \sqrt{(c+e)^2-4b(a+d+f)} } {2b},
\non \\
x_2=\frac{2d+e(1-\beta_q)}{s\beta_q (1+\beta_q)} \hspace*{2.2cm} \ \ ,
\hspace*{1cm}
& & y_{2\pm} =\frac{-d-e \pm \sqrt{(d+e)^2-4f(a+b+c)} } {2(a+b+c)}, \non \\
x_3= -\frac{2d+e(1-\beta_q)}{s\beta_q (1-\beta_q)} \ \  \hspace*{1.9cm} ,
\hspace*{1cm} & & y_{3\pm} =\frac{-d \pm \sqrt{d^2-4af}} {2a}.
\end{eqnarray}

\newpage

\subsection*{Figure Captions}

\renewcommand{\labelenumi}{Fig.\arabic{enumi}}
\begin{enumerate}

\vspace{6mm}
\item
The decays $Z,h,P \rightarrow \glgl$ proceed via 1--loop diagrams with two
quark propagators and one squark propagator (1a), or two (possibly different)
squark propagators and one quark propagator (1b).

\vspace{6mm}
\item
The branching ratio for the decay of the light scalar Higgs boson $h$ of the
MSSM into a pair of light gluinos is shown as a function of the mass of the
lighter stop eigenstate $\tilde{t}_1$, for the ratio of vacuum expectation
values $\tanb=25$ (2a) and 2 (2b), respectively. We have fixed $\msq=400$ GeV,
and $\mu = 200$ GeV for all cases except the dotted curve in fig.~2b, which is
for $\mu=500$ GeV. The curves have been obtained by varying the $A$ parameter
in the region $A<0$, as explained in the text.

\vspace{6mm}
\item
The branching ratio for the decay of the pseudoscalar Higgs boson $P$ of the
MSSM into a pair of light gluinos is shown as a function of $m_{\tilde{t}_1}$,
for $\msq=400$ GeV, $\mu=200$ GeV, $m_P=50$ GeV and three different values of
\tanb\ as indicated. The curves for small \tanb\ terminate well below
the possible maximum of $m_{\tilde{t}_1}$ since here substantial 1--loop
corrections to the scalar Higgs sector are necessary to evade the ALEPH bound
\cite{aleph} on the $ZZh$ coupling; this leads to a lower bound on $|A_t|$ in
these cases.

\vspace{6mm}
\item
The branching ratio for the decay of the $Z$ boson into a pair of light
gluinos is shown as a function of the mass $m_{\tilde{q}_L}$ of $SU(2)$
doublet squarks, for $\mu=200$ GeV and various combinations of $A$ and \tanb.
The solid and dashed curves have been derived using our usual assumption of
equal explicit SUSY breaking masses for $SU(2)$ doublet and singlet squarks,
while the dotted curve is for $m^2_{\tilde{q}_R} = 2 m^2_{\tilde{q}_L}$.

\vspace{6mm}
\item
Scatter plots in the planes spanned by the branching ratios of $h$ and $P$
(5a), $h$ and $Z$ (5b) and $P$ and $Z$ (5c) into pairs of light gluinos. These
plots have been obtained by randomly choosing combinations of input parameters
within the boundaries 50 GeV $\leq \msq \leq$ 500 GeV, 50 GeV $\leq \mu \leq$
1 TeV, 25 GeV $\leq m_P \leq$ 500 GeV, 1.2 $\leq \tanb \leq$ 25 and $-4 \leq A
\leq 4$. Combinations of parameters that violate the bounds on squark masses
and the $ZZh$ coupling discussed in the text have been discarded.

\end{enumerate}


\newpage

\begin{thebibliography}{99}
\frenchspacing
\bibitem{1}
L. Clavelli, Phys. Rev. {\bf D46}, 2112 (1992); L. Clavelli, P.W. Coulter and
K. Yuan, Phys. Rev. {\bf D47}, 1973 (1993); M. Jezabek and J.H. K\"uhn,
Phys. Lett. {\bf B301}, 121 (1993); J. Ellis, D.V. Nanopoulos and D.A. Ross,
Phys. Lett. {\bf B305}, 375 (1993); L. Clavelli and P.W. Coulter, Alabama
Univ. report  UAHEP--941 (1994).

\bibitem{farrar}
G.R. Farrar, Rutgers Univ. report RU--94--35, hep--ph 9407401.

\bibitem{3}
A. De Rujula and R. Petronzio, Nucl. Phys. {\bf B261}, 587;
G. Altarelli, B. Mele and S. Petrarca, Phys. Lett. {\bf B160}, 317 (1985);
V. Barger, S. Jacobs, J. Woodside and K. Hagiwara, Phys. Rev. {\bf D33}, 57
(1986); R.M. Barnett, H.E. Haber and G.L. Kane, Nucl. Phys. {\bf B267}, 625
(1986).

\bibitem{pp}
UA1 collab., C. Albajar et al., Phys. Lett. {\bf B198}, 261 (1987);
UA2 collab., J. Alitti et al., Phys. Lett. {\bf B235}, 363 (1990);
CDF collab., F. Abe et al., Phys. Rev. Lett. {\bf 69}, 3439 (1992).

\bibitem{mssm}
J.F. Gunion and H.E. Haber, Nucl. Phys. {\bf B272}, 1 (1986).

\bibitem{rev}
For a recent review on the phenomenology of Higgs bosons in the
MSSM see, A. Djouadi, Report UdeM-GPP-94--01, to appear in Int. J. Mod. Phys.
A.

\bibitem{valle}
F. de Campos and J.W.F. Valle, Valencia Univ. report FTUV--93--9A,
hep--ph 9311231.

\bibitem{diaz}
J.L. Lopez, D.V. Nanopoulos and X. Wang, Phys. Lett. {\bf B313}, 241 (1993);
M.A. Diaz, Vanderbilt Univ. report VAND--TH--94--7, hep--ph 9404234.

\bibitem{ng}
K. Ng, H. Pois and T.~C.~Yuan, Phys. Rev. {\bf D40}, 1689 (1989).

\bibitem{7}
B.A. Campbell, J.A. Scott and M.K. Sundaresan, Phys. Lett. {\bf B126}, 376
(1983); G.L. Kane and W.B. Rolnick, Nucl. Phys. {\bf B217}, 117 (1983).

\bibitem{norway}
B. Kileng and P. Osland, Bergen Univ. report BERGEN--1994--10,
hep--ph 9407290.

\bibitem{rudaz}
J. Ellis and S. Rudaz, Phys. Lett. {\bf B128}, 248 (1983).

\bibitem{dn1}
M. Drees and M.M. Nojiri, Nucl. Phys. {\bf B369}, 54 (1992).

\bibitem{dn7}
M. Drees and M.M. Nojiri, Phys. Rev. {\bf D49}, 4595 (1994).

\bibitem{corr}
E. Braaten and J.P. Leveille, Phys. Rev. {\bf D22}, 715 (1980); M. Drees and
K.--I. Hikasa, Phys. Lett. {\bf B240}, 455 (1990); A. Djouadi and P. Gambino,
report UdeM-GPP-94--02, to appear in Phys. Rev. D.

\bibitem{l1}
Y. Okada, M. Yamaguchi and T. Yanagida, Prog. Theor. Phys. {\bf 85}, 1 (1991);
H.E. Haber and R. Hempfling, Phys. Rev. Lett. {\bf 66}, 1815; R. Barbieri,
M. Frigeni and F. Caravaglio, Phys. Lett. {\bf B258}, 395 (1991).

\bibitem{l2}
J. Ellis, G. Ridolfi and F. Zwirner, Phys. Lett. {\bf B262}, 477 (1991); M.
Drees and M.M. Nojiri, Phys. Rev. {\bf D45}, 2482 (1992).

\bibitem{aleph}
ALEPH collab., D. Decamp et al., Phys. Lett. {\bf B265}, 475 (1991).

\bibitem{opal}
OPAL collab., R. Akers et al., CERN report PPE--94--104.

\bibitem{fcnc}
B.A. Campbell, Phys. Rev. {\bf D28}, 209 (1983).

\bibitem{DDK} A. Djouadi, M. Drees and H. K\"onig, Phys. Rev. {\bf D48}, 3081
(1993).

\bibitem{cuypers}
F. Cuypers, Phys. Rev. {\bf D49}, R3075 (1994).

\bibitem{lepgl}
R. Munoz--Tapia and W.J. Stirling, Phys. Rev. {\bf D49}, 3763 (1994).

\bibitem{hpp}
V. Barger, M.S. Berger, A.L. Stange and R.J.N. Phillips, Phys. Rev. {\bf D45},
4128 (1992); J.F. Gunion and L. Orr, Phys. Rev. {\bf D46}, 2052 (1992); Z.
Kunszt and F. Zwirner, Nucl. Phys. {\bf B385}, 3 (1992).

\end{thebibliography}
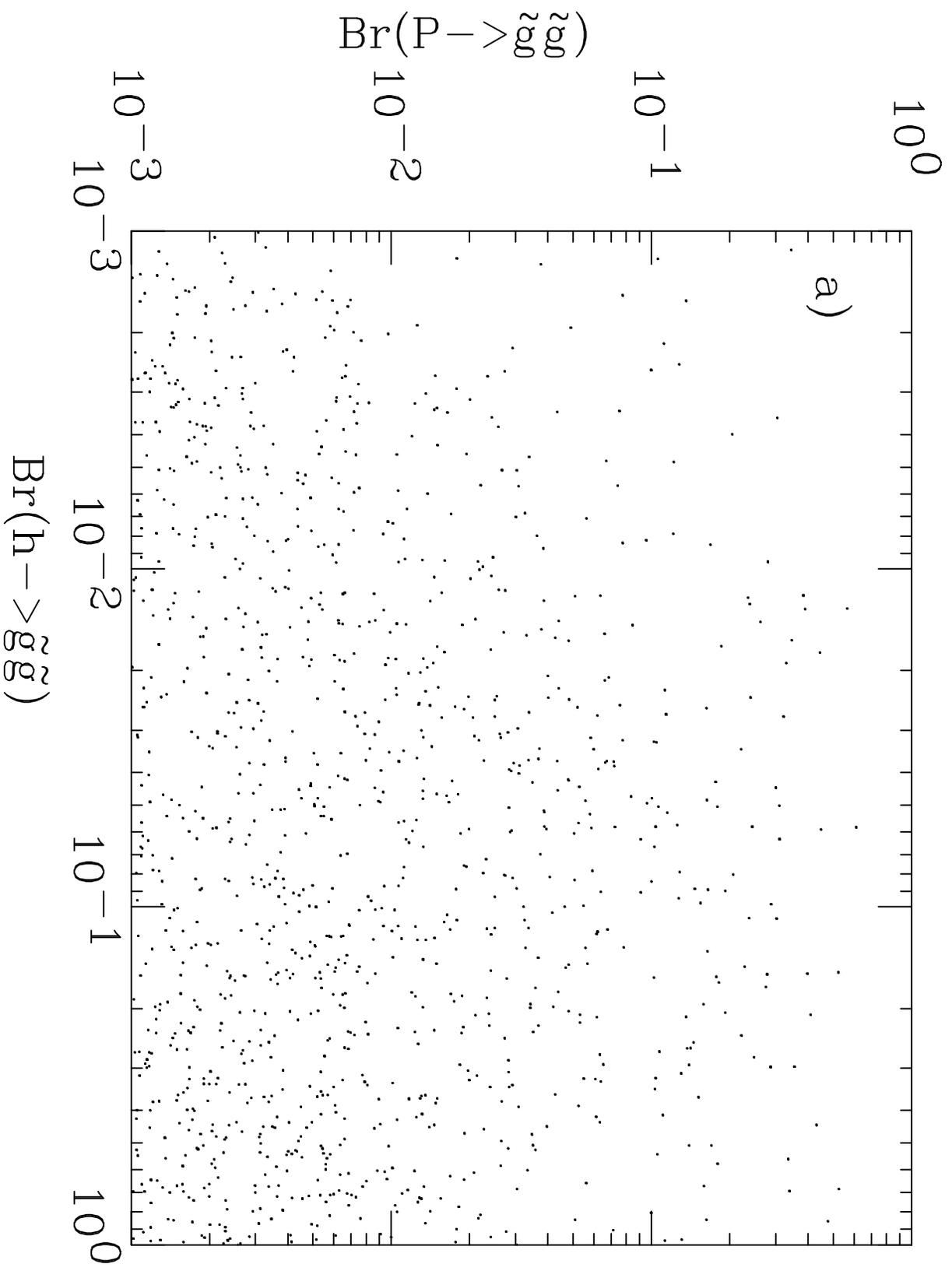
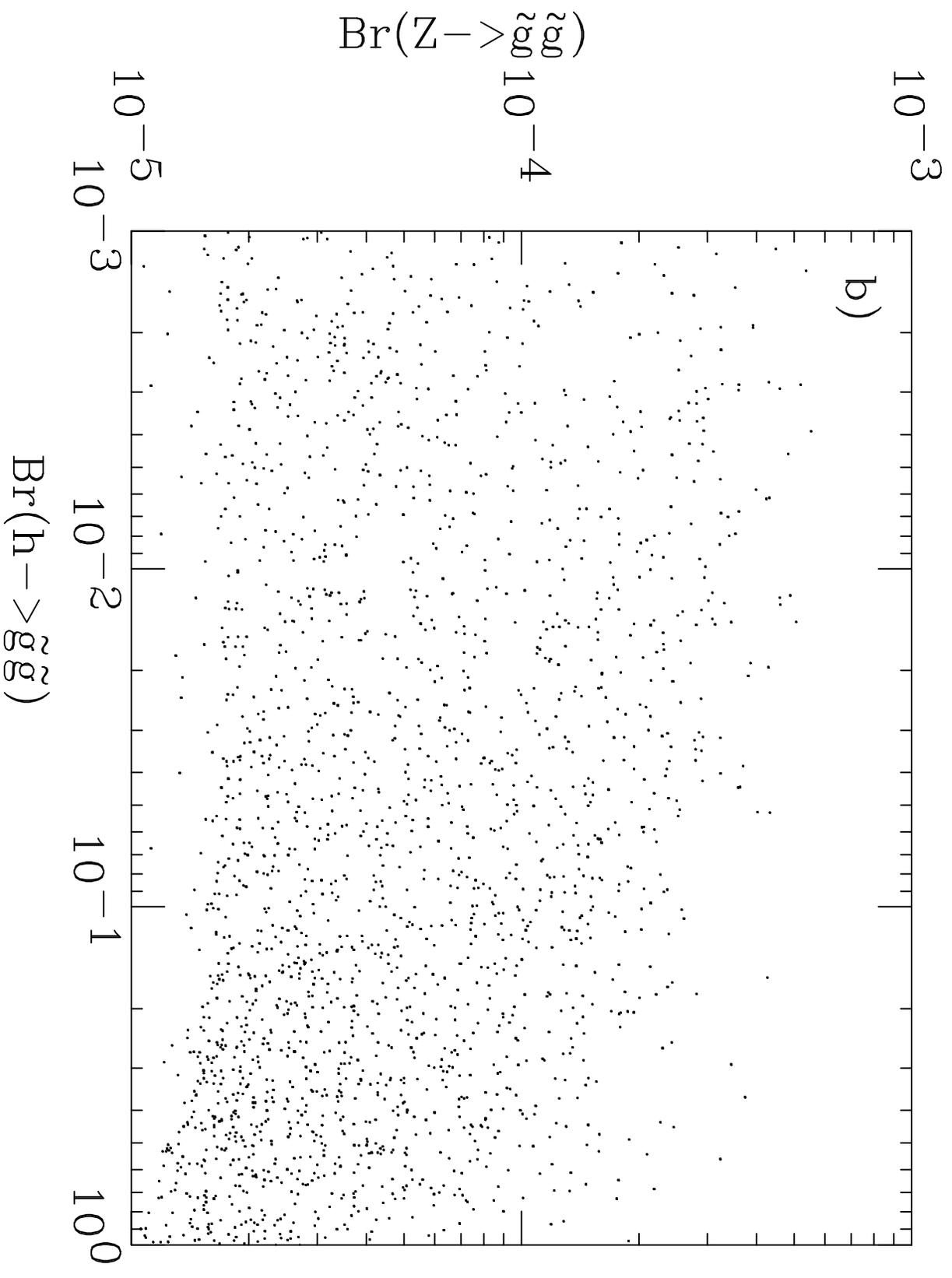
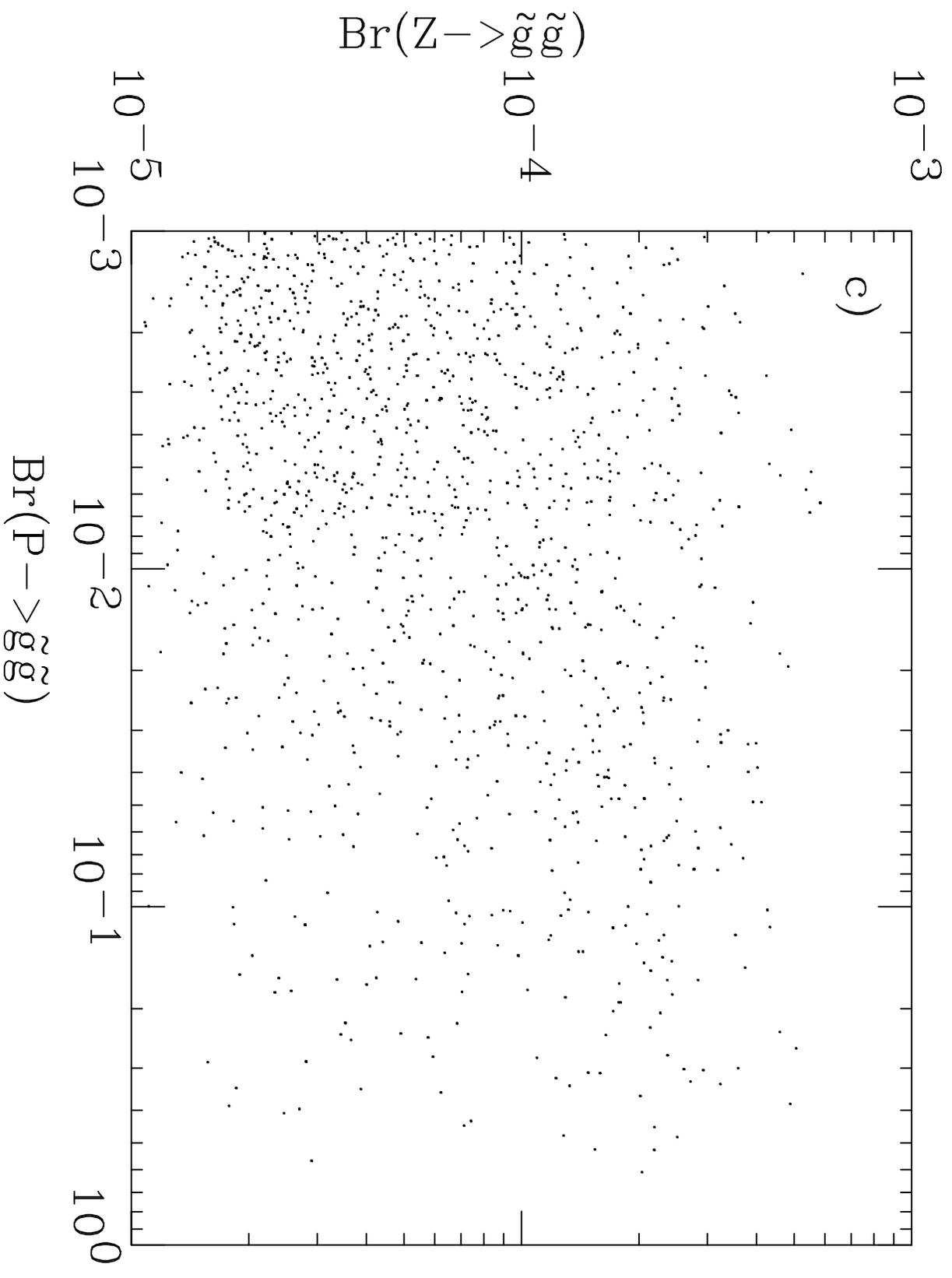
\end{document}